\documentclass[aps,prc,twocolumn,floatfix,superscriptaddress]{revtex4}
\usepackage{epsfig}
\usepackage{amsmath}
\usepackage{color}
\usepackage{graphicx}
\definecolor{blue}{rgb}{0.05, 0.05, 0.5}
\def \beq{\begin{equation}}
\def \eeq{\end{equation}}
\def \beqa{\begin{eqnarray}}
\def \eeqa{\end{eqnarray}}

\newcommand{\nn}{\nonumber}

\usepackage{xspace}

\begin{document}
\title{Effects of initial state nucleon shadowing on the elliptic flow of thermal photons}
\author{Pingal Dasgupta}
\email{pingaldg@vecc.gov.in}
\affiliation{Variable Energy Cyclotron Centre, 1/AF, Bidhan Nagar, Kolkata-700064, India}
\affiliation{Homi Bhabha National Institute, Training School Complex, Anushakti Nagar, Mumbai 400085, India}
\author{Rupa Chatterjee}
\email{rupa@vecc.gov.in}
\affiliation{Variable Energy Cyclotron Centre, 1/AF, Bidhan Nagar, Kolkata-700064, India}
\affiliation{Homi Bhabha National Institute, Training School Complex, Anushakti Nagar, Mumbai 400085, India}
\author{Sushant K. Singh}
\email{sushantsk@vecc.gov.in}
\affiliation{Variable Energy Cyclotron Centre, 1/AF, Bidhan Nagar, Kolkata-700064, India}
\affiliation{Homi Bhabha National Institute, Training School Complex, Anushakti Nagar, Mumbai 400085, India}
\author{Jan-e Alam}
\email{jane@vecc.gov.in}
\affiliation{Variable Energy Cyclotron Centre, 1/AF, Bidhan Nagar, Kolkata-700064, India}
\affiliation{Homi Bhabha National Institute, Training School Complex, Anushakti Nagar, Mumbai 400085, India}
\begin{abstract}
Recently the effect of nucleon shadowing on the Monte-Carlo Glauber initial condition was 
studied and its role on the centrality dependence of elliptic flow ($v_2$) and fluctuations in initial
eccentricity for different colliding nuclei were explored. It was found that the results with 
shadowing effects are closer to the QCD based dynamical model as well as to the experimental data.
Inspired by this outcome, in this work we study the transverse momentum ($p_T$)
spectra and elliptic flow of thermal photons for Au+Au 
collisions at RHIC and Pb+Pb collisions at LHC by incorporating the shadowing effects
in deducing the initial energy density profile required to solve the relativistic hydrodynamical equations. 
We find that the thermal photon spectra remain almost unaltered, however, the elliptic 
flow of photon is  found to be  enhanced significantly due to shadowing effects. 

\end{abstract}

\maketitle

\section{Introduction} 
The primary goal of heavy ion collision programmes at Relativistic Heavy Ion Collider (RHIC)
and Large Hadron Collider (LHC) is to produce a new state of thermalized matter called quark gluon plasma
(QGP) - where the properties of the system are not governed by hadrons but by the quarks and gluons.
Collision of nuclei will create charged particles  either
in  the form of hadrons or partons depending on the magnitude of the collision
energy. Electromagnetic interaction among these charged particles will
inevitably lead to the production of photons with mean free path much larger than the size of the
system formed in such collisions. As a result  photons can bring the information of
the production point very efficiently without getting affected by the 
secondary interactions and therefore, 
considered as one of the efficient probes for the 
detection of QGP~\cite{phot}. This has led to huge theoretical
~\cite{dks,cgale,rupa1,cshen,amonnai,lmclerran,fliu,gbasar,ktuchin,zakharov,gvujanovic} and 
experimental~\cite{wa98, phenix, alice} efforts
to study the mechanism of photon production in heavy ion collisions at relativistic energies (HICRE). 
Recent data from PHENIX~\cite{phenix} 
and ALICE~\cite{alice} Collaborations  at the RHIC and LHC respectively have reported 
excess of direct photons over the scaled yield from proton-proton collisions
in the transverse momentum domain, $p_T <$ 4 GeV. This excess  is attributed to the thermal radiation 
from QGP and hadronic matter.  However, in spite of a large number of detailed studies by
several authors~\cite{sryu,jfp,vhees,vhees2,olinyuk} a simultaneous explanation of the data on $p_T$ 
spectra and differential elliptic flow  of photons is still lacking
both at RHIC~\cite{phenix_v2}  and at the LHC~\cite{alice_v2} energies, the lack of this 
explanation has been dubbed as ``direct photon puzzle''~\cite{cern}.

The QGP produced in heavy ion collisions at RHIC and LHC energies evolve in space and time and revert to hadrons at transition
temperature, $T_c$.  
In such a scenario  photons  will be produced through various processes at different
stages of the evolution populating different domains of $p_T$.  These are
broadly categorized as: (i) prompt photons originating 
from the interactions of the partons of the colliding  nuclei which will populate
the high $p_T$ domain;
(ii) thermal productions - from the interactions of thermal partons as well as
from thermal hadrons  - will occupy low and intermediate $p_T$  and finally (iii) from the decays of the long lived
(compared to strong interaction time scale) mesons. Photons from  (i) are non-thermal
therefore, are affected neither by the temperature nor by the flow. For a given
collision energy this contribution could be subtracted out from the data by using pQCD results.
Photon spectra from proton+proton collision may be used as a benchmark to 
validate theoretical results. Photons from the decays of hadrons have been subtracted 
out in the data presented by ALICE and PHENIX Collaborations~\cite{alice, phenix}. 
Photons from thermal source, (ii) are sensitive to temperature and flow, therefore, 
this is the component of the spectra that needs urgent attention to address the ``direct photon puzzle''. 

Elliptic flow is considered as one of the fundamental observables of collectivity of the 
system produced in HICRE. Thermal photons originating from expanding QGP
along with prompt photons explain the data on the photon spectra both for Au+Au collisions at RHIC and 
Pb+Pb collisions at LHC energies in the region $p_T >$ 2 GeV~\cite{chre1,chre2}. 
However, the large discrepancies between theoretical results and the 
data on thermal photon's $v_2$  remains a  puzzle.
Several studies such as calculations based on sophisticated event-by-event viscous hydrodynamic
model, studies incorporating pre-equilibrium contributions and including the effects of high initial magnetic field etc. are unable to explain the  data on $v_2$ of photons till date. 

With the increase in the precision of experimental measurements, it becomes imperative
for theoretical calculations to include finer physical effects. One such effect is
shadowing of nucleons deep inside the colliding nuclei by the nucleons at the 
front during the process of collisions.  
In an effort to understand the correlation between the multiplicity and eccentricity,  
effect of shadowing was included in Refs.~\cite{janeda1,janeda2} in the Monte-Carlo Glauber model 
to deduce the initial energy density profile which  is required as an input to the relativistic hydrodynamical
equations.  Solutions of relativistic hydrodynamical  equations
are used to understand the development of collectivity in the system. 
It was observed that the inclusion of shadowing in the Monte-Carlo (MC) Glauber  
model increases the elliptic flow significantly compared to the result obtained 
from the conventional MC Glauber initial condition. In this paper, we explore  the effects of 
nucleon shadowing on the elliptic flow of photons  at RHIC and LHC energies.

We set the goal of the paper at this point. In this work our aim is not to 
reproduce the experimental data on $v_2$ and thermal spectra of photons or
to solve the ``direct photon puzzle" but to estimate  $v_2$ by including 
the effects of shadowing that a nucleon deep inside the colliding
nucleus is subjected to. 
In this context we evaluate $v_2$ and thermal spectra with and without 
shadowing effects by  fixing the value of charged hadron multiplicity 
to its experimental value.

The paper is organized as follows. In the next section we will discuss the MC Glauber
model with and without shadowing effects. Production mechanisms of photons from QGP and HM 
have been briefly discussed in section III with appropriate references for details. 
The effects of shadowing on
the hydrodynamic evolution and elliptic flow of thermal photon spectra  have been presented
in section IV. Section V is devoted to summary and conclusions.  

\section{Initial conditions}
In the conventional MC Glauber model, all the nucleons are given equal 
weightage for energy deposition i.e. a nucleon undergoing multiple collisions 
will deposit the same amount of energy in each collision. In Ref.~\cite{janeda1} 
it is argued that the nucleons located deep inside the nucleus are eclipsed or shadowed by the 
nucleons in the front. Therefore, the contribution of a participating nucleon to energy deposition 
will crucially depend on its position in the colliding nuclei. This is accomplished by introducing a weight 
factor $S(n,\,\lambda)$ in the initial state as follows:
\begin{equation}
S(n,\,\lambda)\,=\,\exp(-n\lambda) 
\label{sheqn}
\end{equation}
where, $S(n,\lambda)$ accounts for shadowing on a participant due to $n$ other nucleons in the same  
nucleus which are in front and  conceal it partially (see Refs.~\cite{janeda1, janeda2} for more detail). 
In the present work we took $\lambda = 0.12$ and $0.08$ at RHIC and LHC respectively~\cite{janeda1,janeda2}. 
We call the MC Glauber initial condition without the shadowing effect as MCG and the initial condition with the shadowing as shMCG.

A MCG model with standard two-parameter Woods-Saxon  nuclear density profile is used to randomly 
distribute the nucleons into the two colliding nuclei. Two nucleons from different nuclei are assumed to collide 
when the relation $d^2 < \frac{\sigma_{NN}}{\pi}$ is satisfied where, $d$ is the transverse distance between the 
colliding nucleons and $\sigma_{NN}$ is the inelastic nucleon-nucleon cross-section. We take $\sigma_{NN}$ as 42 mb and
64 mb for 200A GeV Au+Au collision at RHIC and 2.76A TeV Pb+Pb collision at LHC respectively. 

For the collision of two nuclei, we assume the beam axis to be along $z$-direction and impact 
parameter to be along $x$-direction for a particular event. The plane spanned by the $x$- and $y$-axes 
is the transverse plane. The initial entropy distribution, $s(x,y)$, is obtained by first locating $(x,y)$-coordinates 
of participants and binary collisions which are treated as sources of energy deposition, and then taking a weighted sum 
over all the sources as explained in Eq.~\ref{imple_eqn} below. The coordinates of a binary collision is taken as the 
average of the coordinates of two colliding nucleons. Hence, sources are distributed randomly in the transverse plane 
and each source receives a different weight for energy deposition depending on whether the source is a participant or 
a binary collision. 
For MCG, a participant is given a weight $(1-\nu)$ and a binary collision is given a weight of $\nu$, where 
$\nu$ is some constant which is obtained by fitting with the experimental data of charged hadron multiplicity distribution ($dN_{\rm ch}/d\eta$). The entropy density is then obtained using the following expression,
 \begin{equation}
\label{imple_eqn}
 s(x,y)=K\sum _{i=1}^{N_s} w_i(\Theta _i)\,f_i(x,y)
 \end{equation}
where $\Theta _i$ is a binary variable used to label the source and which decides the type of weight 
to be given making $w_i$ a function of $\Theta _i$ which is denoted by $w_i(\Theta _i)$. 
We take $w_i=\nu$ and $(1-\nu)$ for  
$\Theta _i=0$ and  $\Theta _i=1$  respectively.
In Eq.~\ref{imple_eqn}, $N_s$ denotes the total number of sources and $f_i(x,y)$ is the normalized distribution given by,
 \begin{equation}
 \label{normal_dist_eq}
 f_i(x,y)=\frac{1}{2\pi \sigma ^2}\, e^{ -\frac{(x-x_i)^2+(y-y_i)^2}{2\sigma ^2}}. 
\end{equation}
Note that in the limit $\sigma \rightarrow 0$, Eq.\ref{imple_eqn} reduces to the two component formula :
\begin{equation}
 s(x,y)=K\left[ \, \nu \, n_{\text{coll}}(x,y)+(1-\nu)\, n_{\text{part}}(x,y)\, \right]
\end{equation}
where $n_{coll}$ and $n_{part}$ are number of collisions and number of participants respectively.
The form of entropy density in shMCG is same as in Eq.~\ref{imple_eqn} except that the weights $w_i$ are different. 
For $\Theta _i=0$, $w_i=\nu\, S_i^c$ and for $\Theta _i=1$, $w_i=(1-\nu)S_i^p$ where $S_i^p$ and $S_i^c$ are the shadowing weights. 
$S^p_i$ is given by Eq.~\ref{sheqn}, denotes weight factor when a participant nucleon is subjected to due to shadowing.  
To obtain the shadowing weights, we follow Ref~\cite{janeda1}. Each nucleon is first assigned a weight depending on how many other 
nucleons are in front. $S_i^p$ is then the weight of the corresponding wounded nucleon and $S_i^c$ is the product of 
weights of the nucleons undergoing the collision.  In Eq.~\ref{normal_dist_eq}, $\sigma$ is a free parameter, taken as 
$\sigma=$ 0.4 fm~\cite{hannu,chre1} for both MCG and shMCG. 

We take the values of initial thermalization times for RHIC and LHC as $\tau_0=$ 0.17 fm/$c$ and  $ 0.14$ fm/c respectively
from the EKRT mini-jet saturation model~\cite{ekrt}. 
In the present work, (2+1) dimensional inviscid relativistic hydrodynamic model 
has been used to study the space-time evolution of the matter produced in HICRE~\cite{hannu}. 
Same hydrodynamical model  with MCG initial condition has been used earlier to study spectra and 
anisotropic flow of hadrons and photons~\cite{hannu, chre1, chre2}. We modify the initial conditions to include the shadowing effect. 

The initial entropy density profile required as input to hydrodynamical calculation 
is constructed for both the cases (MCG and shMCG) by taking initial state average of $N(=10000)$ 
random events (where events obey the distribution $dN/db\propto b$, $b$ is impact parameter)
within the particular centrality class as follows :
\begin{equation}
\label{eq.ncollex}
s(x,y) = \frac{1}{N} \sum_{i=1}^{N} s_{i}(x,y)
\end{equation} 
where, $s_{i}(x,y)$ characterizes the entropy distribution of the $i^{th}$ event produced according to the Eq.~\ref{imple_eqn}.
The initial flow velocity component, $v_x$ and $v_y$ are taken as zero here.
\section{Photons from thermal QGP and hadrons}
Contributions from the QGP matter to the thermal photon spectra due to
annihilation ($q$$\bar{q}$$\rightarrow$$g$$\gamma$) and
the QCD Compton ($q(\bar{q})g\rightarrow q(\bar{q})\gamma$)
processes to the order $\alpha_s\alpha$ was estimated in~\cite{kapusta,bair}
by  using hard thermal loop (HTL) approximation~\cite{braaten}.
Later, it was shown that photons from the processes~\cite{aurenche1}:
$g$$q$$\rightarrow$$g$$q$$\gamma$,
$q$$q$$\rightarrow$$q$$q$$\gamma$,
$q$$q$$\bar{q}$$\rightarrow$$q$$\gamma$
and $g$$q$$\bar{q}$$\rightarrow$$g$$\gamma$
contribute also to the same order as $O(\alpha\alpha_s)$.
The complete calculation of emission rate from QGP to order $\alpha_s$
has been performed by resumming ladder diagrams in the effective  field
theory~\cite{arnold,nlo_thermal}.
In the present work, the rate of production of thermal photons has been  taken
from~\cite{arnold}.  The $T$ dependence of the strong coupling, $\alpha_s$
has been taken from~\cite{zantow}.

For the photon production from hadronic matter
an exhaustive set of hadronic reactions and the radiative
decays of resonances are considered. The rate has been
taken from ~\cite{trg} which includes the effects of
the hadronic form factor (see also ~\cite{we1,we2,we3}).

The $p_T$ distribution of photons is obtained by integrating the 
temperature dependent emission rates ($R=EdN/d^3pd^4x$) 
over the entire space-time evolution history - from the initial 
thermalization time to the final freeze-out state of the fireball via intermediary quark-hadron 
transition as :
\beqa
E \frac{dN}{d^3p}&=&\int d^4x\,\lbrace R_Q\left(E^*(x),T(x)\right)\Theta(T-T_c)\nn\\
&&+R_H\left(E^*(x),T(x)\right)\Theta(T_c-T)\Theta(T-T_F)\rbrace
\label{emrate}
\eeqa
where, $T(x)$ is the local temperature and $E^* (x)$ = $p^\mu u_\mu (x)$ is the energy in
the comoving frame, $p^\mu$ is the four-momentum of the photons and $u_\mu$ is the 
local four-velocity of the flow field, $R_Q (R_H)$ is the emission rate of photons from QGP (hadronic system).
$T$ and $u^\mu$ are obtained from
the solution of hydrodynamical equations.

The elliptic flow parameter $v_2$  is  calculated by using the relation :
\begin{equation}
v_2 (p_T) \ = \ \langle {\rm cos}(2\phi) \rangle \ = \ \frac{\int_0^{2\pi} \,
d\phi\,{\rm cos}(2\phi)\,\frac{dN}{p_T dp_T dy d\phi}}{\int_0^{2\pi}\,d\phi\,\frac{dN}{p_Tdp_Tdy d\phi}}.
\label{v2flow}
\end{equation}
The temperature at freeze-out is taken as 160 MeV which reproduces the measured $p_T$ 
spectra of charged pions at RHIC and LHC energies. The value of quark-hadron transition
temperature is taken as 170 MeV and the lattice QCD based EoS is taken from~\cite{eos}
to solve the hydrodynamical equations.

\section{Results}
We study the effects of shadowing by considering the following two cases:
\\
\\
Case-I : First we consider the case where $\nu=0$. 
This gives the wounded nucleon profile. We study the difference in the evolution scenario 
with and without shadowing effects.
In the following, we calculate averages of hydrodynamic quantities by using the following relation: 
\begin{equation}
\langle f \rangle=\frac{\int \mathrm{d}x \mathrm{d}y \ f(x,y)\  \epsilon(x,y,\tau)}{\int \mathrm{d}x \mathrm{d}y \  \epsilon(x,y,\tau)} \,
\end{equation}
where, $\epsilon(x,y,\tau)$ is the energy density at $(x,y)$ at proper
time, $\tau$  obtained by solving the hydrodynamic equations.
\begin{figure}
\centerline{\includegraphics*[width=8.1cm,clip=true]{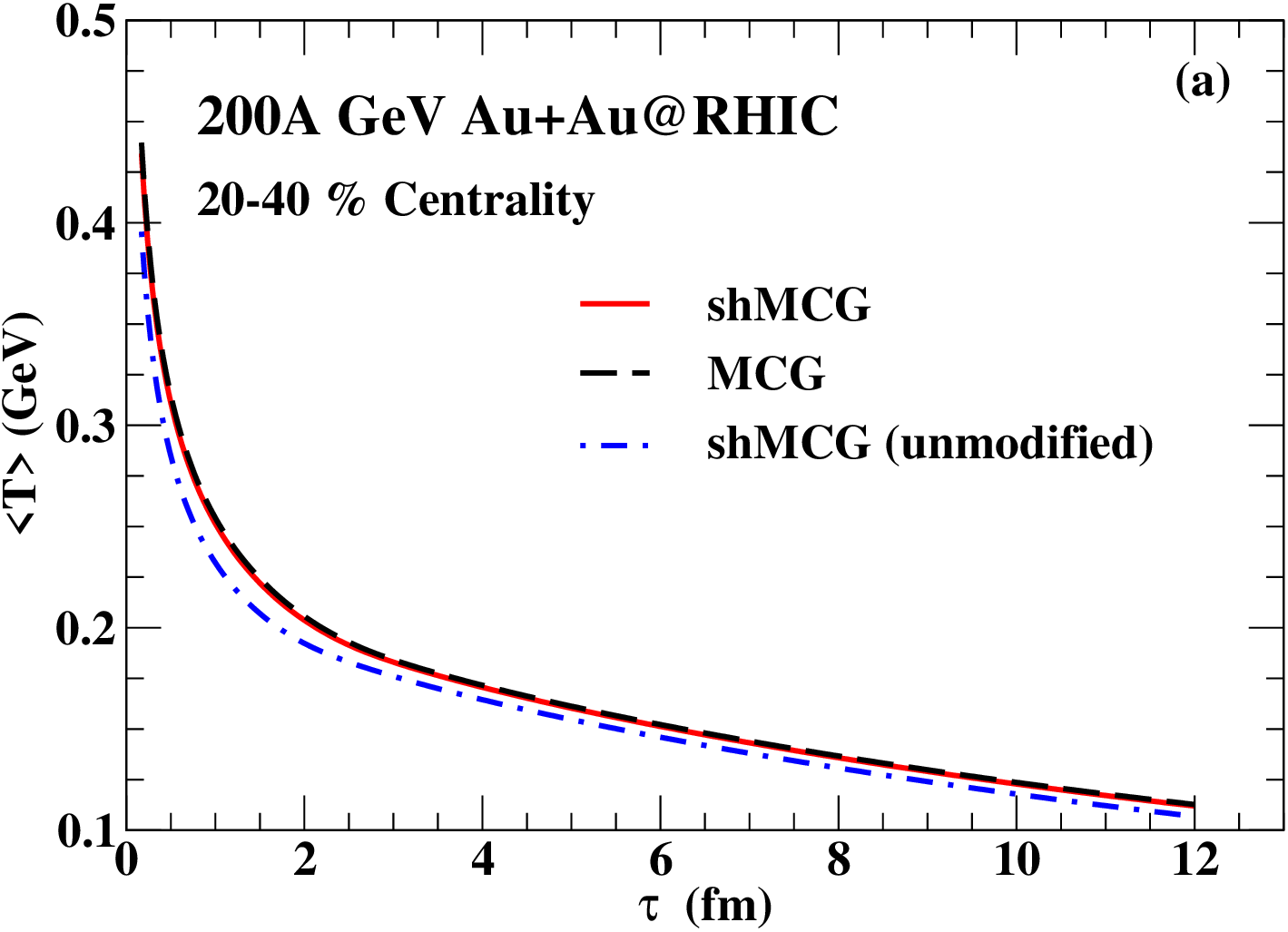}}
\centerline{\includegraphics*[width=8.1cm,clip=true]{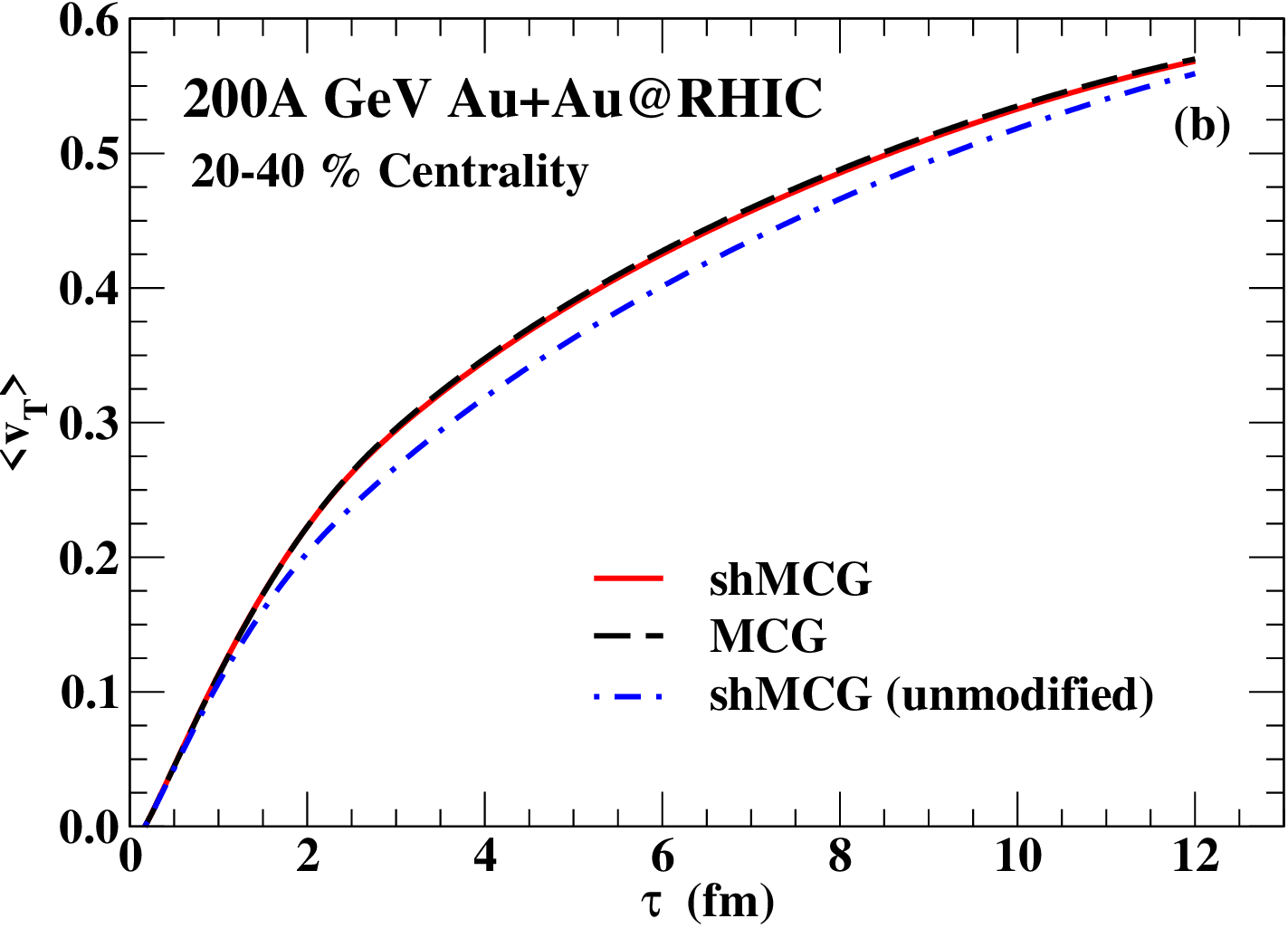}}
\caption{(Color online) (a) Time evolution of average temperature and (b) transverse flow velocity with MCG and 
shMCG initial conditions have been depicted for $\nu=0$.}
\label{evol_wn}
\end{figure}

 We construct the MCG initial condition by taking $K=102 \text{ fm}^{-1}$ in Eq.~\ref{imple_eqn}. 
The MCG initial condition has been used extensively  earlier to calculate photon production at RHIC and LHC energies~\cite{chre1,chre2}. 
The time evolution of average temperature, $\langle T\rangle$ and average transverse flow velocity, 
$\langle v_T\rangle$ for MCG initial condition for 200A GeV Au+Au collisions at RHIC in 
$20-40\%$ centrality bin 
are shown by black dashed lines in Fig.~\ref{evol_wn}(a) 
and Fig.~\ref{evol_wn}(b) respectively. 
The inclusion of initial state shadowing, \emph{i.e.}, $\lambda \neq 0$, in the MCG initial condition and 
without changing $K$ (from now on we call this as unmodified shMCG) results in smaller charged hadron multiplicity or total entropy. 
This leads to smaller average temperature (as shown by blue dotted lines in Fig.~\ref{evol_wn}(a)) and pressure, 
which in turn results in smaller average transverse flow velocity (shown by blue dotted lines in Fig.~\ref{evol_wn}(b)). 

In order to obtain the initial condition for shMCG, the normalization factor $K$ (Eq.~\ref{imple_eqn}) 
is tuned to reproduce the same final $dN_{\rm ch}/d\eta$ as MCG initial condition. We obtain $K=140 \text{ fm}^{-1}$ 
for shMCG. The time evolution of $\langle T \rangle$ and $\langle v_T \rangle$  for the shMCG initial condition are shown by red solid lines in Fig.~\ref{evol_wn}(a) and Fig.~\ref{evol_wn}(b) respectively. 
It is to be noticed that the differences in $\langle T \rangle$ and $\langle v_T \rangle $ for MCG and shMCG are  insignificant
when the model parameters in shMCG are adjusted to reproduce the same charged hadron multiplicity as in MCG.

\begin{figure}
\centerline{\includegraphics*[width=8.1cm,clip=true]{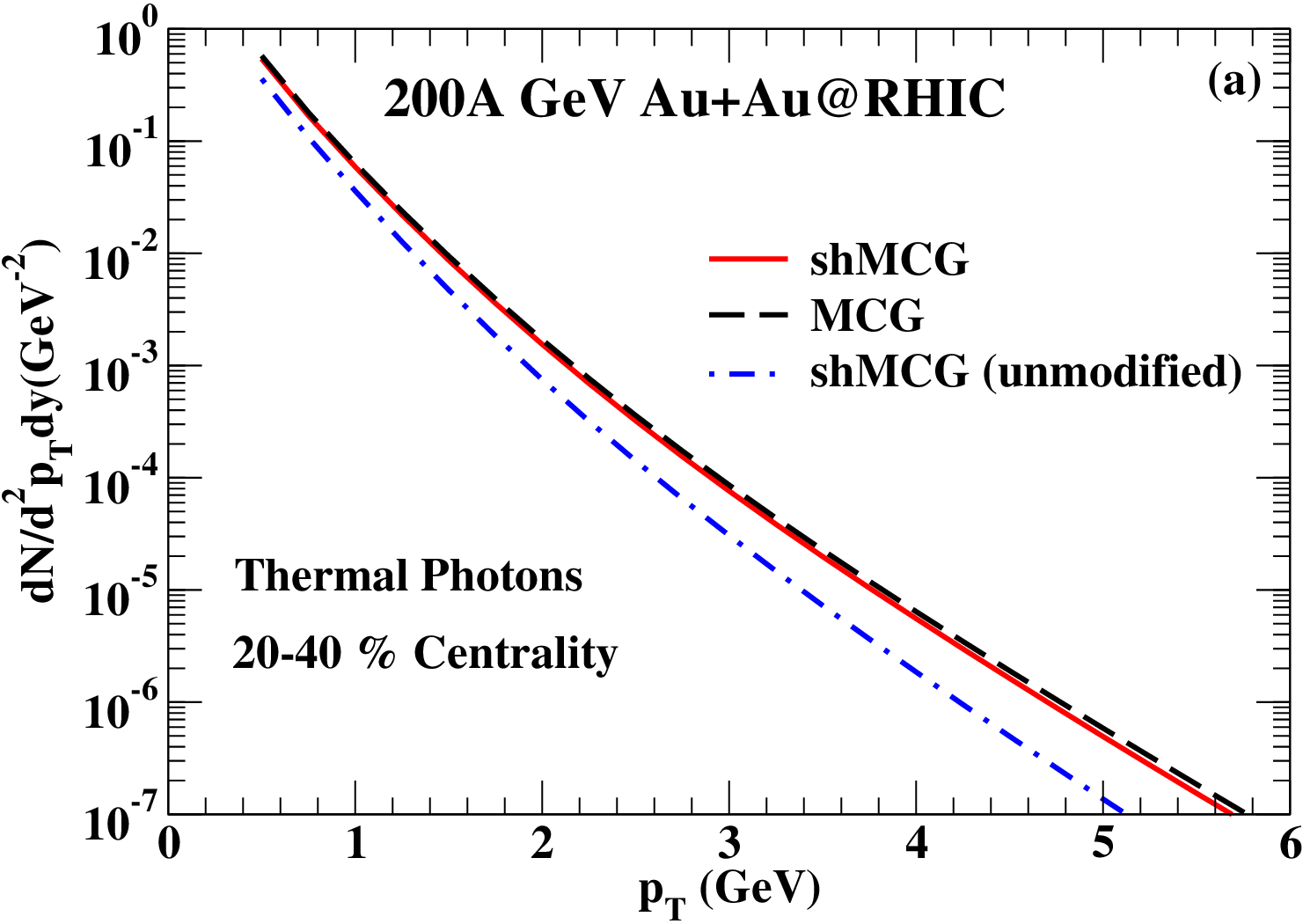}}
\centerline{\includegraphics*[width=8.1cm,clip=true]{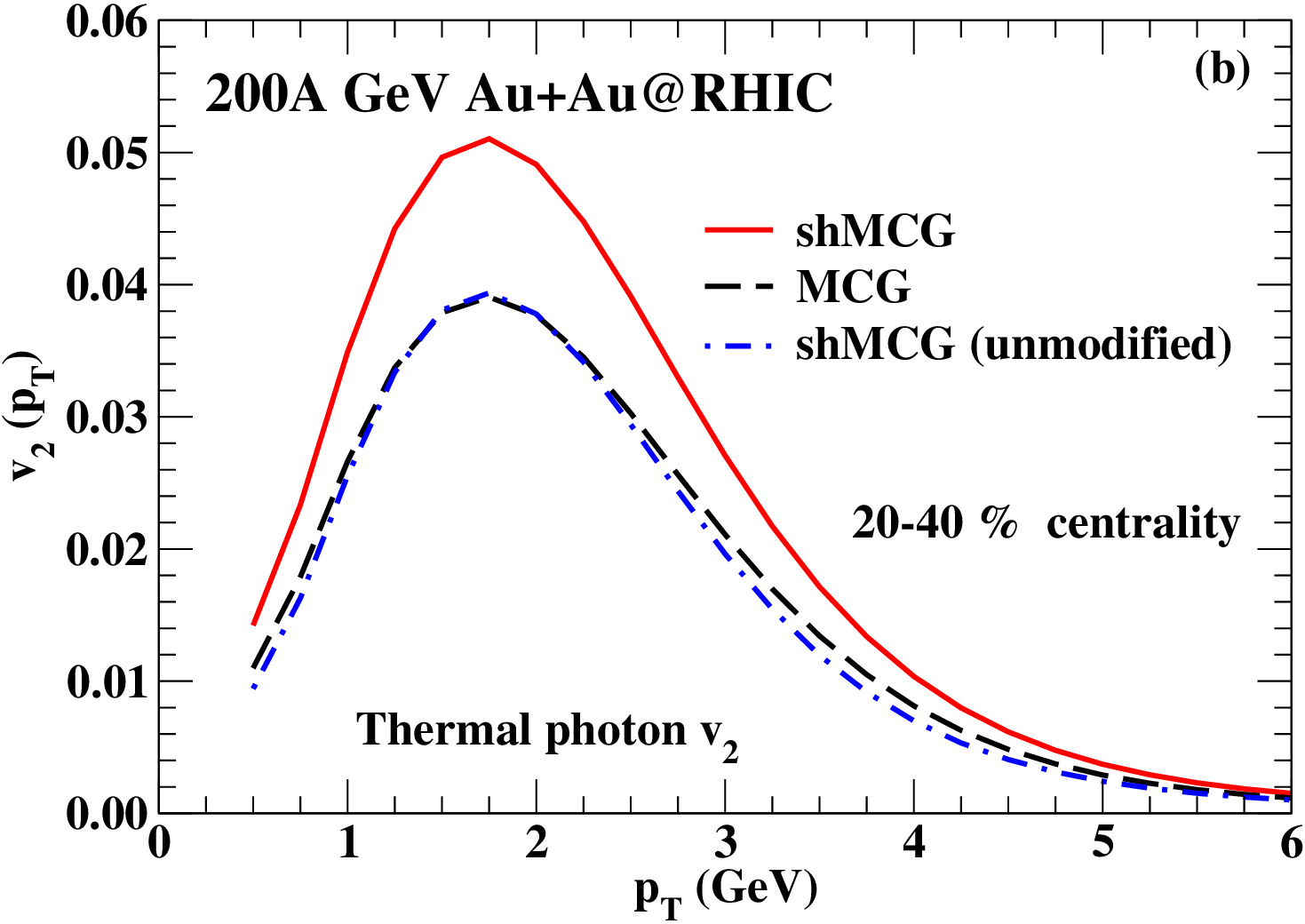}}
\caption{(Color online) (a) Thermal photon spectra and (b) elliptic flow from MCG and shMCG initial conditions for wounded profile at RHIC 200A GeV Au+Au collisions in 20--40\% centrality bin.}
\label{wn_phot}
\end{figure}

Next  we display our results on the $p_T$ spectra of photons. Fig.~\ref{wn_phot}(a) shows the thermal 
photon spectra 
with MCG and shMCG initial conditions for 20--40\% central Au+Au collisions 
at RHIC. For given production rate of photons with similar space-time evolution 
scenario, the spectra for the two initial conditions (MCG and shMCG) are found to be close to 
each other. For both the cases, photon spectra are dominated by radiation from the QGP  for 
$p_T >$ 1 GeV and only in the low $p_T$ ( $<$ 1 GeV) region we see significant contribution 
from the hadronic matter. However, for unmodified shMCG the spectra is found to be suppressed. 

The elliptic flow of thermal photons calculated using MCG and shMCG initial conditions are 
shown in Fig.~\ref{wn_phot}(b). As shown in earlier studies~\cite{cfhs,chre2}, 
due to the competing contributions from the quark matter and hadronic matter (HM) states, the  $v_2$ of photons shows different nature compared to that of hadrons. 

The  differential elliptic flow of photons is small at large $p_T$ as these
are mostly emitted from the (early)  QGP state where the flow has not developed fully. 
As we approach from high $p_T$, the $v_2(p_T)$ increases towards smaller values of $p_T$, reaches maximum 
around 1.5 -- 2.0 GeV 
and then drops as $p_T$ is reduced further. Such a variation of $v_2(p_T)$ results from the two competing
profiles of elliptic flow originating from QGP and hadronic matter states (see later).

We see a significant increase in the elliptic flow for the shMCG initial condition compared to the MCG initial condition. The value of $v_2(p_T)$ at the peak is about $32\%$ larger in the shMCG case. This may be qualitatively explained as follows. The inclusion of shadowing in the initial condition affects the nucleons situated in the interior of nucleus more strongly than those at the boundary~\cite{janeda1}. This means that the shadowing effects are less prominent at the ends of the major axis of elliptic overlap zone in the transverse plane where the nucleons from the boundary of the two nuclei are involved. Whereas, the shadowing effects are more prominent at the ends of the minor axis where nucleons from the interior of one of the nuclei with large shadowing effects collide with the nucleons from the boundary of the other nucleus. This results in smaller effective length of the 
minor axis in shMCG than MCG, as a consequence the pressure gradient in shMCG is larger resulting in larger elliptic flow. 
Interestingly, one can see that the $v_2$ from the unmodified shMCG initial condition (in which shadowing is included but $K$ is not fixed to reproduce the final particle multiplicity) is close to the $v_2$ from MCG initial condition even when shadowing 
introduces more spatial anisotropy because of the lower pressure (originating from smaller
hadronic multiplicity or entropy density) in case of unmodified shMCG 
scenario (Fig.~\ref{evol_wn}(b)).

\begin{figure}
\centerline{\includegraphics*[width=8.1cm,clip=true]{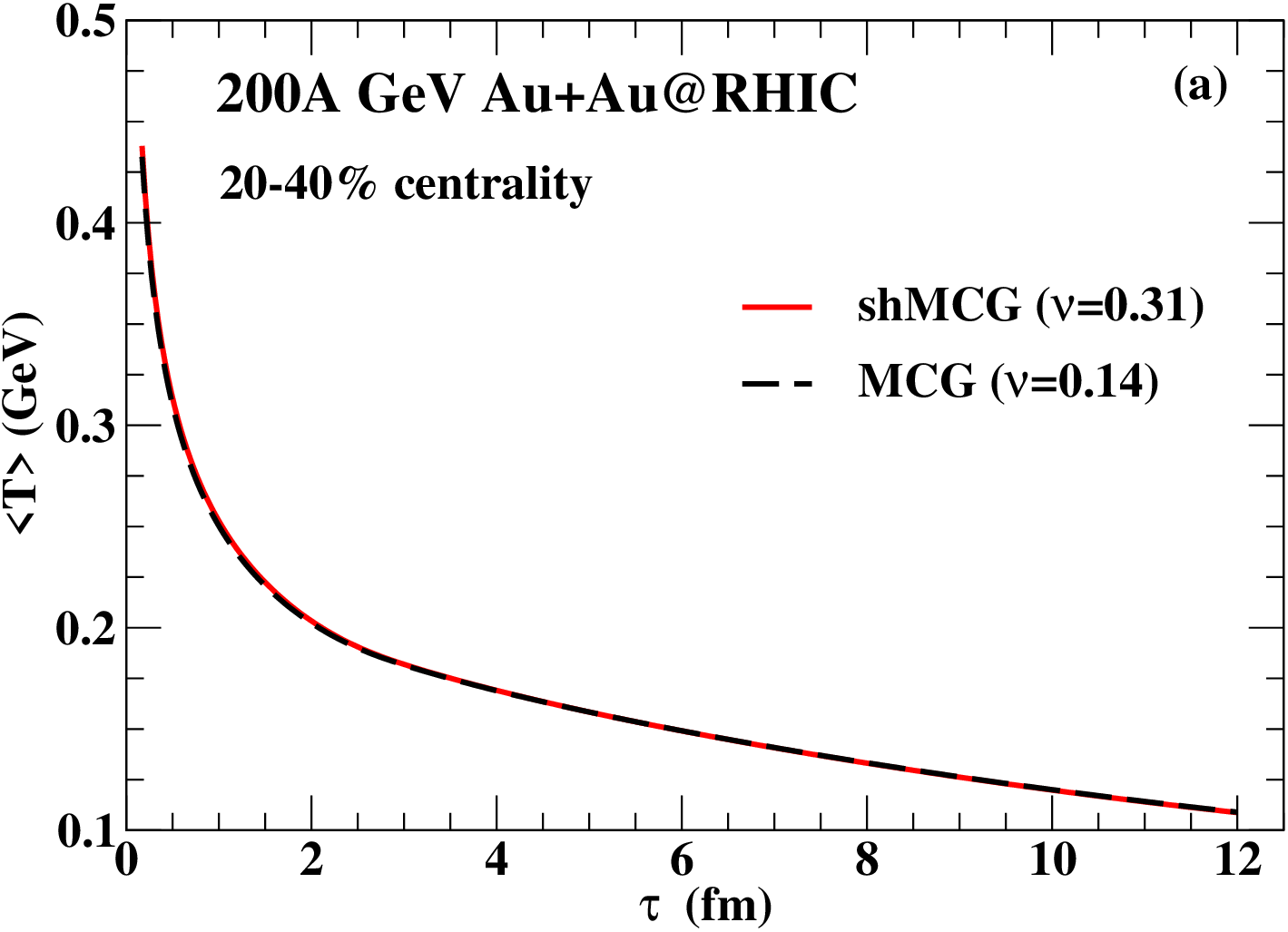}}
\centerline{\includegraphics*[width=8.1cm,clip=true]{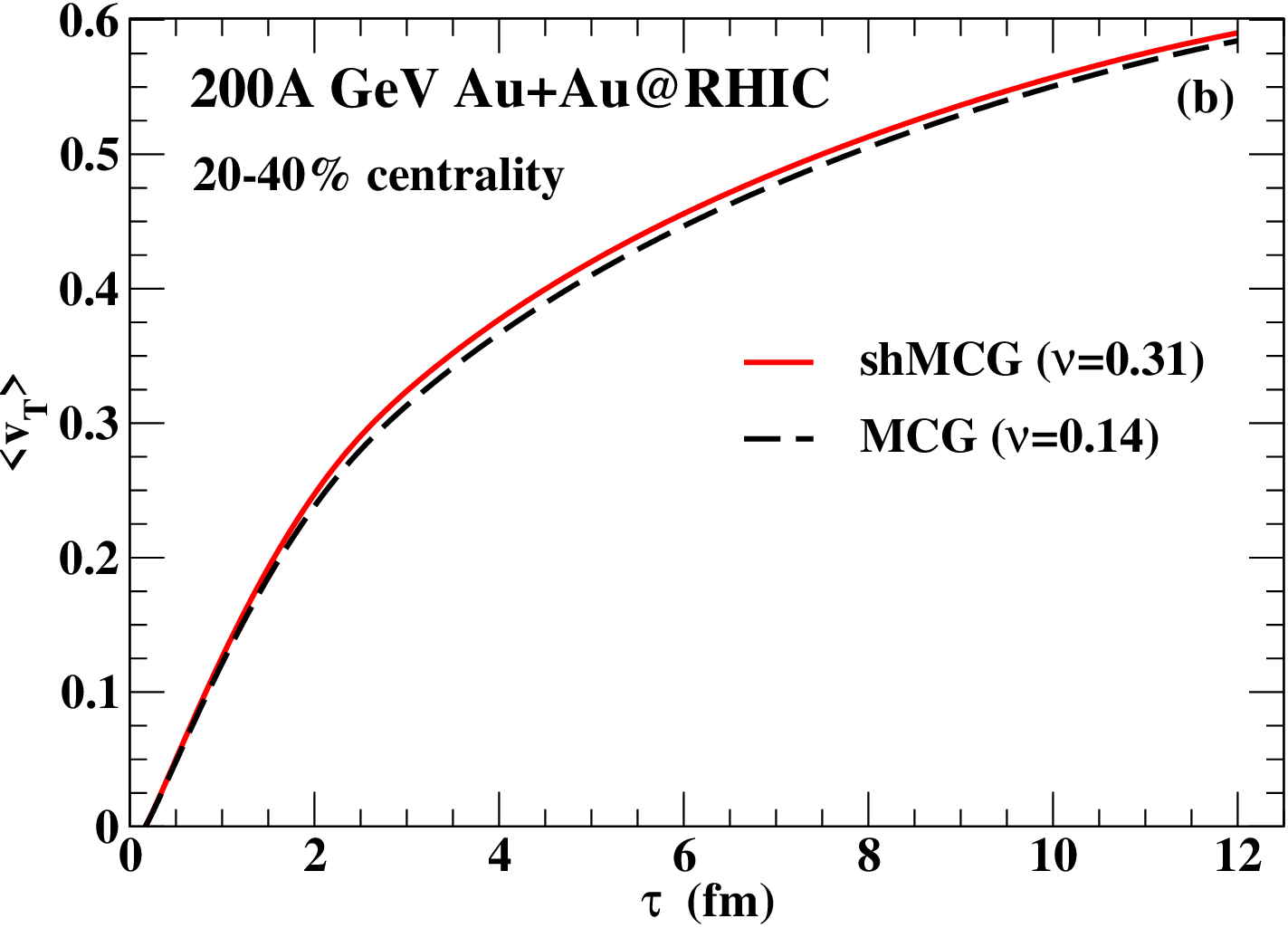}}
\caption{(Color online) (a) Time evolution of average transverse flow 
velocity and (b) average temperature considering MCG and shMCG initial states for two component model
($\nu\neq 0$).}
\label{evol}
\end{figure}

In the present study we consider initial state averaged smooth density distribution and 
show that even though  the photon spectra remain unaltered, the effect of shadowing on the elliptic 
flow  is significantly large.  It has been  shown earlier that event-by-event fluctuating 
initial conditions increase $v_2$ significantly in the region $p_T >$ 2 GeV compared to the $v_2$ calculated 
from a smooth initial state profile~\cite{rupa1}. Thus, result from calculation based on 
event-by-event fluctuating initial condition with shadowing  
is expected to enhance the $v_2$ even more. This  would reduce the difference between the experimental 
data and theoretical results and would help in resolving  
the ``direct photon puzzle''. In addition,  
the triangular flow of photons ($v_3$) which originates from the initial state fluctuations only, is also expected to be 
larger using shMCG initial condition. These aspects of study is postponed for  future~\cite{shadow_ebye_phot_v2}.

\begin{figure}
\centerline{\includegraphics*[width=8.1cm,clip=true]{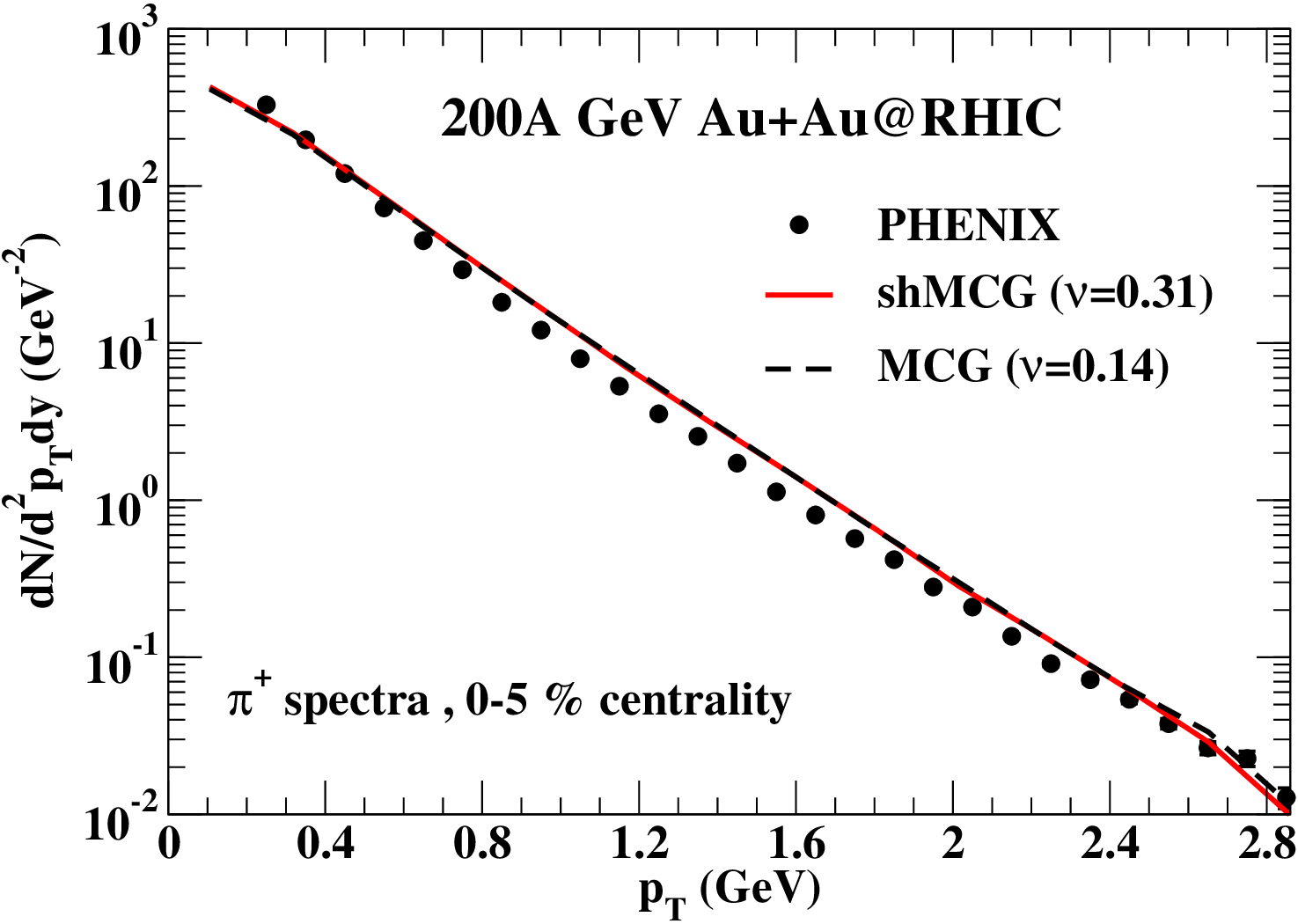}}
\caption{(Color online) The  thermal pion spectra at 
RHIC with two types of initial condition (see text) and comparison with PHENIX data~\cite{Adler:2003cb}.   
}
\label{pion_rhic}
\end{figure}

Case-II :  Next we consider the case $\nu\neq 0$. The nucleon shadowing affects $N_{\rm coll}$ more strongly than $N_{\rm part}$ as shown in Ref.~\cite{janeda1}. Therefore, we consider Au+Au collisions at RHIC where the initial entropy density is taken as proportional to a 
linear combination of $N_{\rm coll}$ and $N_{\rm part}$. We take $\nu=0.14$, $K=80 \text{ fm}^{-1}$ for the MCG initial state and $\nu=0.31$ and 
$K=110 \text{ fm}^{-1}$ for shMCG. It may be noted here that different values of $K$ and $\nu$ in 
shMCG are required to reproduce the same charged hadron multiplicity measured. The transverse momentum spectra of $\pi^+$ has been evaluated 
including the feed-down from higher resonance decays at the freeze-out surface  by using Cooper-Frye~\cite{cooper} formula. The result has been contrasted with experimental data from RHIC~\cite{Adler:2003cb}. We find that the $p_T$ spectra for $\pi^+$ for MCG and shMCG are similar and both are close to experimental data  (Fig.~\ref{pion_rhic}). 

\begin{figure}
\centerline{\includegraphics*[width=8.1cm,clip=true]{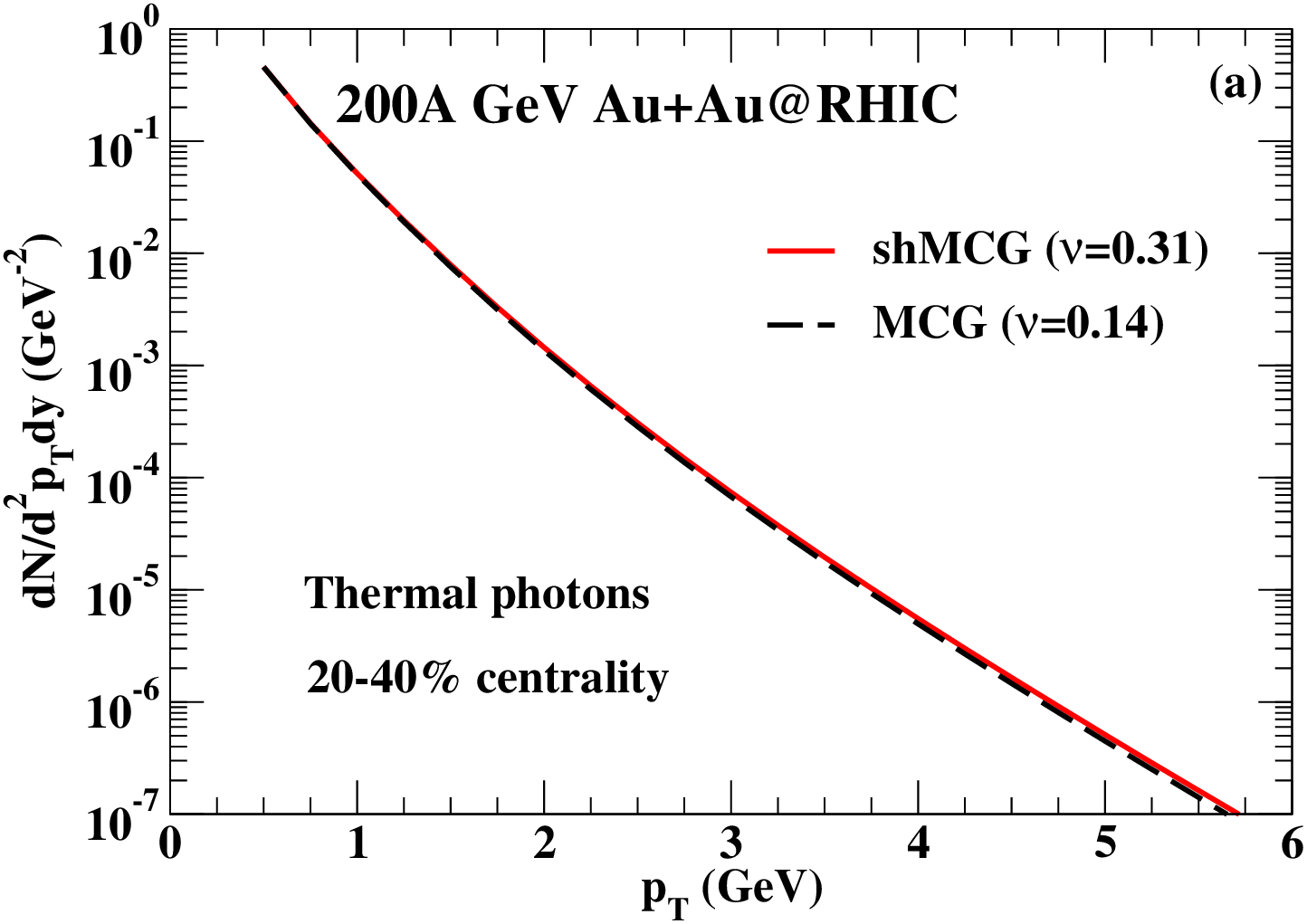}}
\centerline{\includegraphics*[width=8.1cm,clip=true]{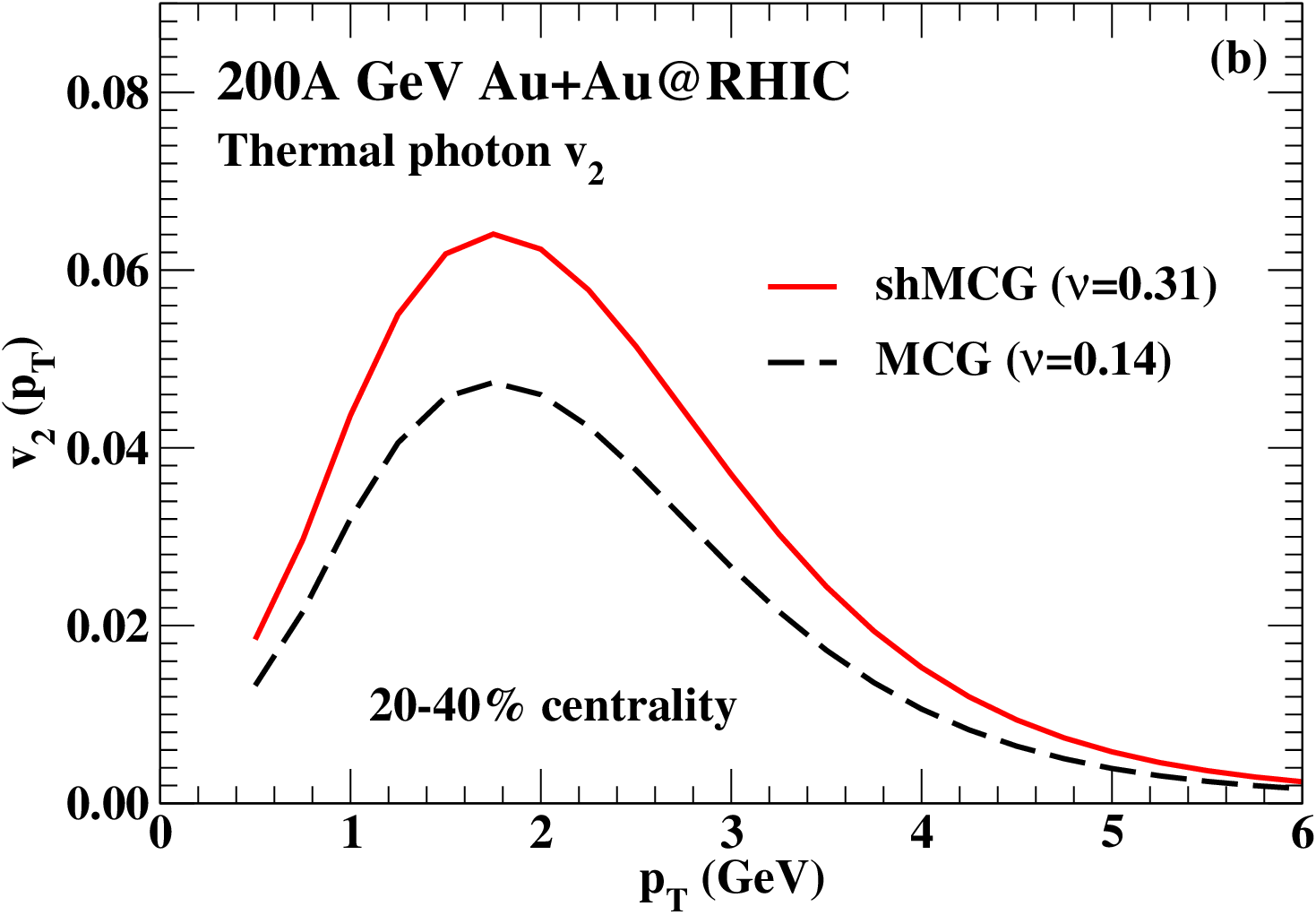}}
\caption{(Color online) (a) Thermal photon spectra and (b) elliptic flow considering MCG and shMCG 
initial conditions for 200A GeV Au+Au collisions at RHIC and 20--40\% centrality 
bin for two component model.}
\label{spec_v2_alpha}
\end{figure}
\begin{figure}
\centerline{\includegraphics*[width=8.1cm,clip=true]{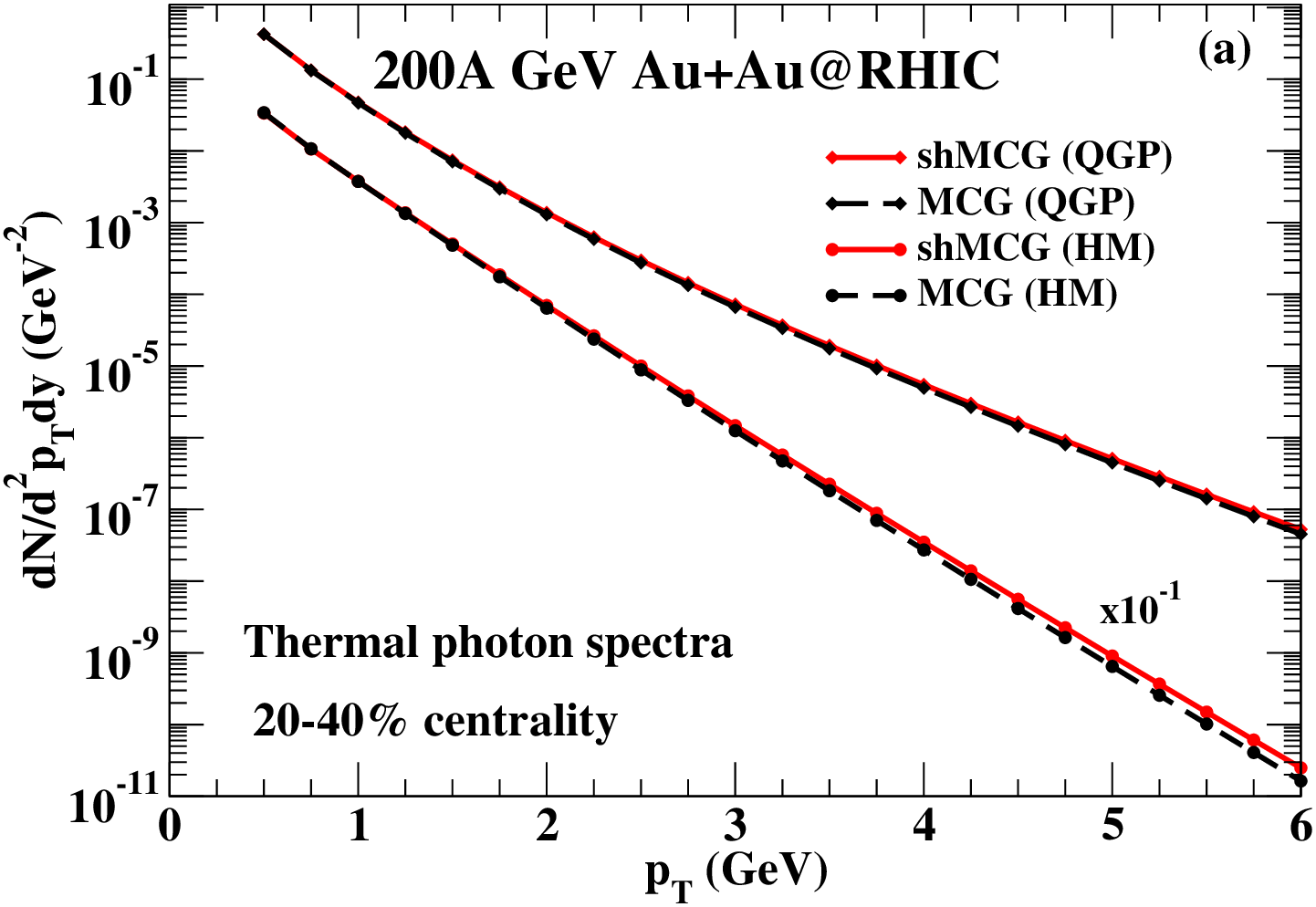}}
\centerline{\includegraphics*[width=8.1cm,clip=true]{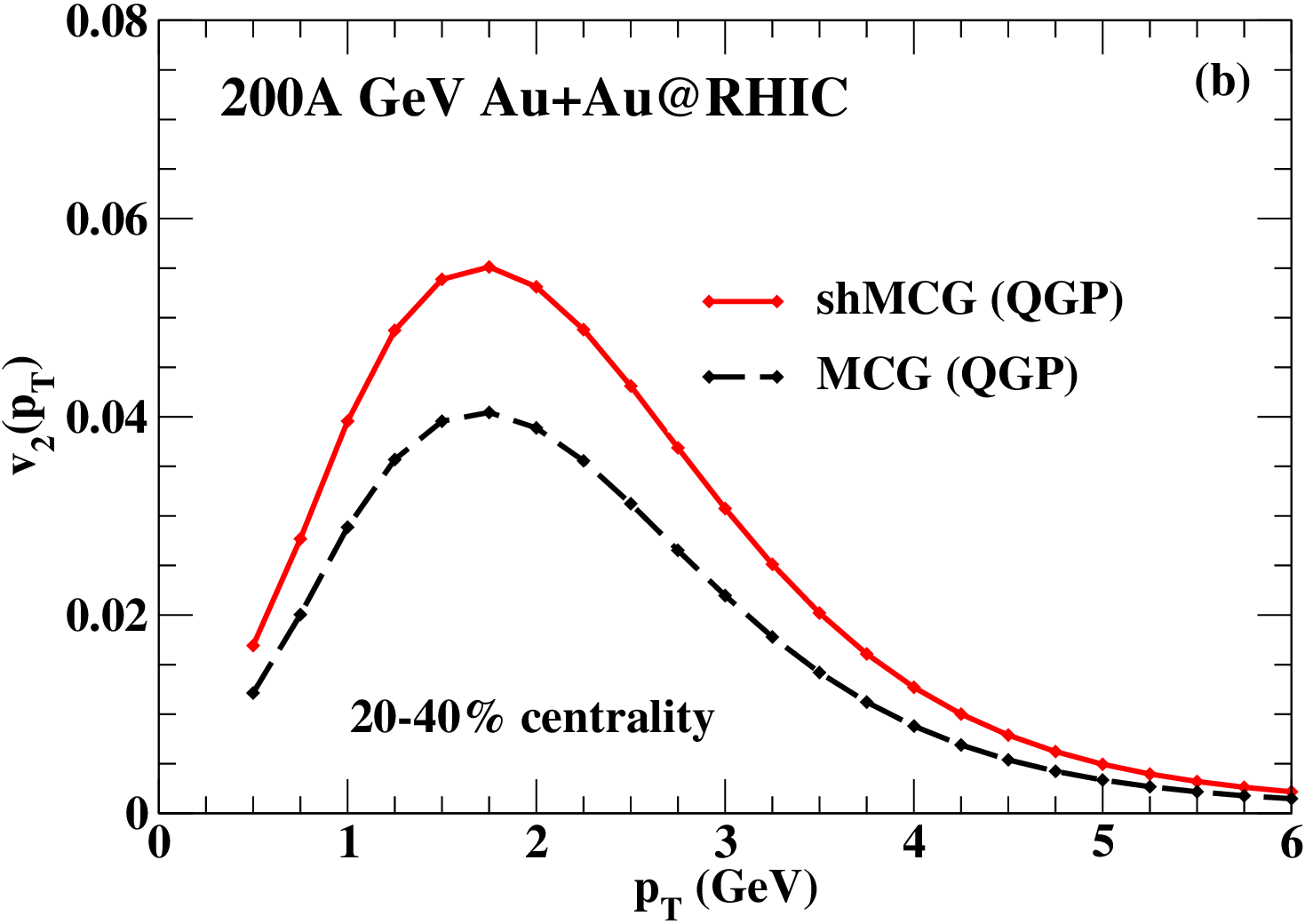}}
\centerline{\includegraphics*[width=8.1cm,clip=true]{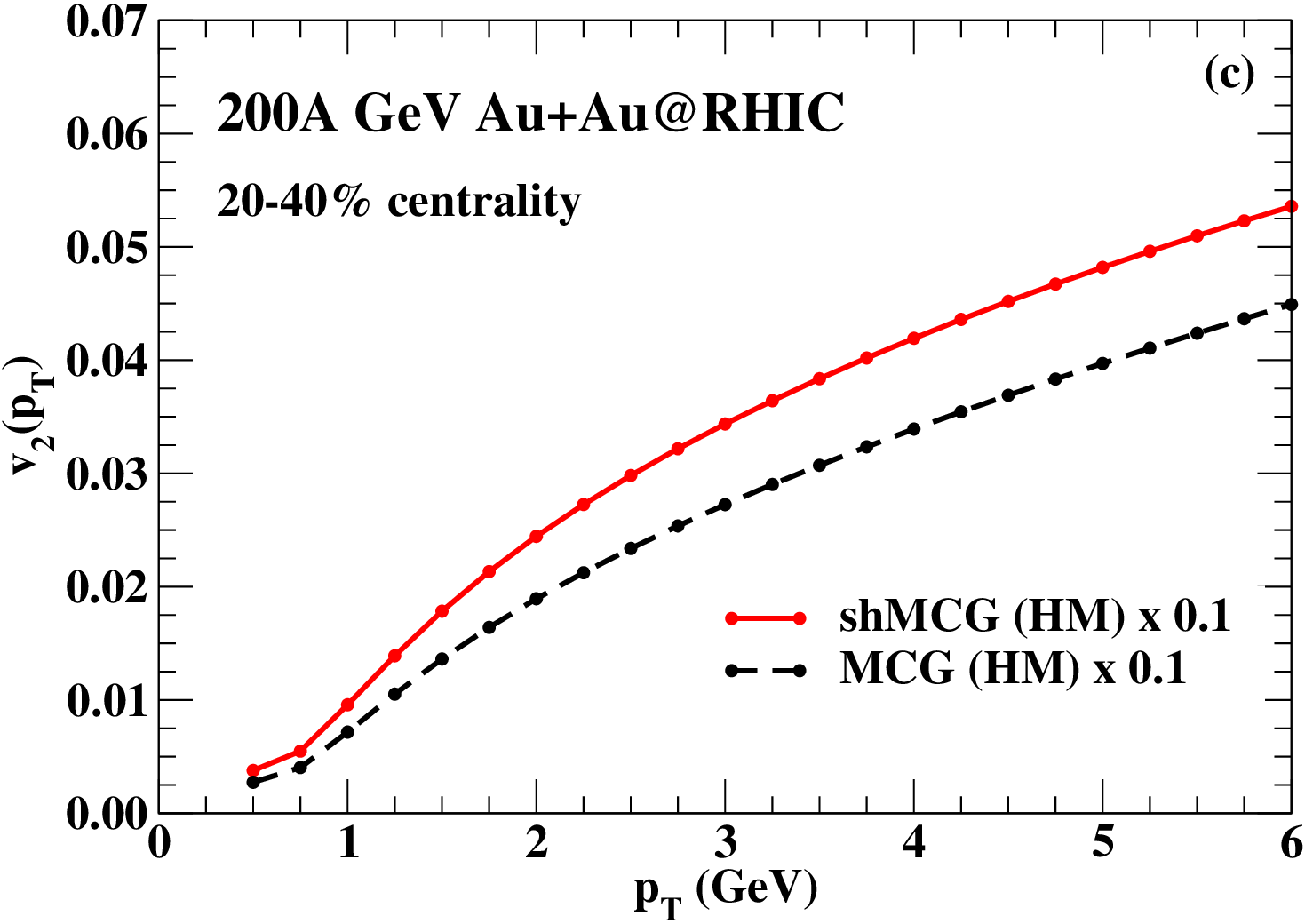}}
\caption{(Color online) (a) Thermal photon spectra and elliptic flow from (b) 
QGP and (c) HM  separately from MCG and 
shMCG initial conditions.}
\label{QGP_HM}
\end{figure}

The time evolution of $\langle T\rangle$  and $\langle v_T\rangle$ for two initial 
conditions (MCG and shMCG) are shown in Fig.~\ref{evol}.  The $\langle T\rangle$ for
the two initial conditions are almost on top of each other whereas, the $\langle v_T\rangle$ is 
found to be  marginally larger for shMCG initial condition. 

For $\nu\neq 0$, successive collisions are allowed as opposed to the case
with $\nu=0$. Therefore, amount of energy deposition is different for
$\nu=0$ and $\nu\neq 0$ as large shadowing will take place in the latter case.  
The pressure gradient in shMCG with $\nu\neq 0$ is larger compared
to the case where $\nu=0$. As a consequence the $\langle v_T\rangle$ in shMCG with $\nu\neq 0$ is slightly larger compared to
the case where $\nu=0$.

The spectra and elliptic flow of thermal photons for $\nu\neq 0$ are shown in 
Fig.~\ref{spec_v2_alpha}(a) and Fig.~\ref{spec_v2_alpha}(b) respectively. 
The nature of the $p_T$ spectra and elliptic flow  is found to be similar to the  results 
from wounded nucleon profile. As shown in the figure, the $p_T$ spectra  for  MCG and 
shMCG initial conditions are again found to be close to each other. 
However, the elliptic flow is slightly larger for two component initial conditions compared to the $v_2$ from 
corresponding single component ($\nu=0$) model. The  peak value of $v_2(p_T)$ for shMCG initial condition is 
about 36\% larger than the $v_2(p_T)$ calculated using the MCG initial condition. 
As explained before the effects of shadowing in the region near the ends of the major axis are 
smaller than region near ends of the minor axis of the elliptic overlap zone, which effectively reduces the minor to major axis ratio in 
shMCG compared to MCG resulting in larger elliptic flow in the shMCG scenario.

In order to understand the effect of initial state shadowing on the photon elliptic flow better, 
we plot the spectra and elliptic flow  from individual QGP and HM  separately 
in Fig.~\ref{QGP_HM}. The thermal photon spectra as expected from both QGP and HM  are close to each 
other for the two different initial conditions as shown in Fig.~\ref{QGP_HM}(a). The effect of initial state shadowing is found to be more pronounced for $v_2$ of photons from QGP (Fig.~\ref{QGP_HM}(b)) compared to $v_2$ from HM  (Fig.~\ref{QGP_HM}(c)) in the region $p_T\sim$ 1.5 to 2.0  GeV (where the increase in total photon $v_2$  due to initial state shadowing is maximum, Fig.~\ref{spec_v2_alpha}). The  $v_2(p_T)$ is about $50\%$
larger in this region  for shMCG  compared to the MCG initial condition in the QGP, whereas the increase  due to shMCG  in
HM is about 30\%.

Next we consider  Pb+Pb collision at 2.76A TeV energy at the LHC.
We evaluate the  $p_T$ spectra of thermal photons  and elliptic flow using MCG and shMCG initial conditions.

\begin{figure}
\centerline{\includegraphics*[width=8.1cm,clip=true]{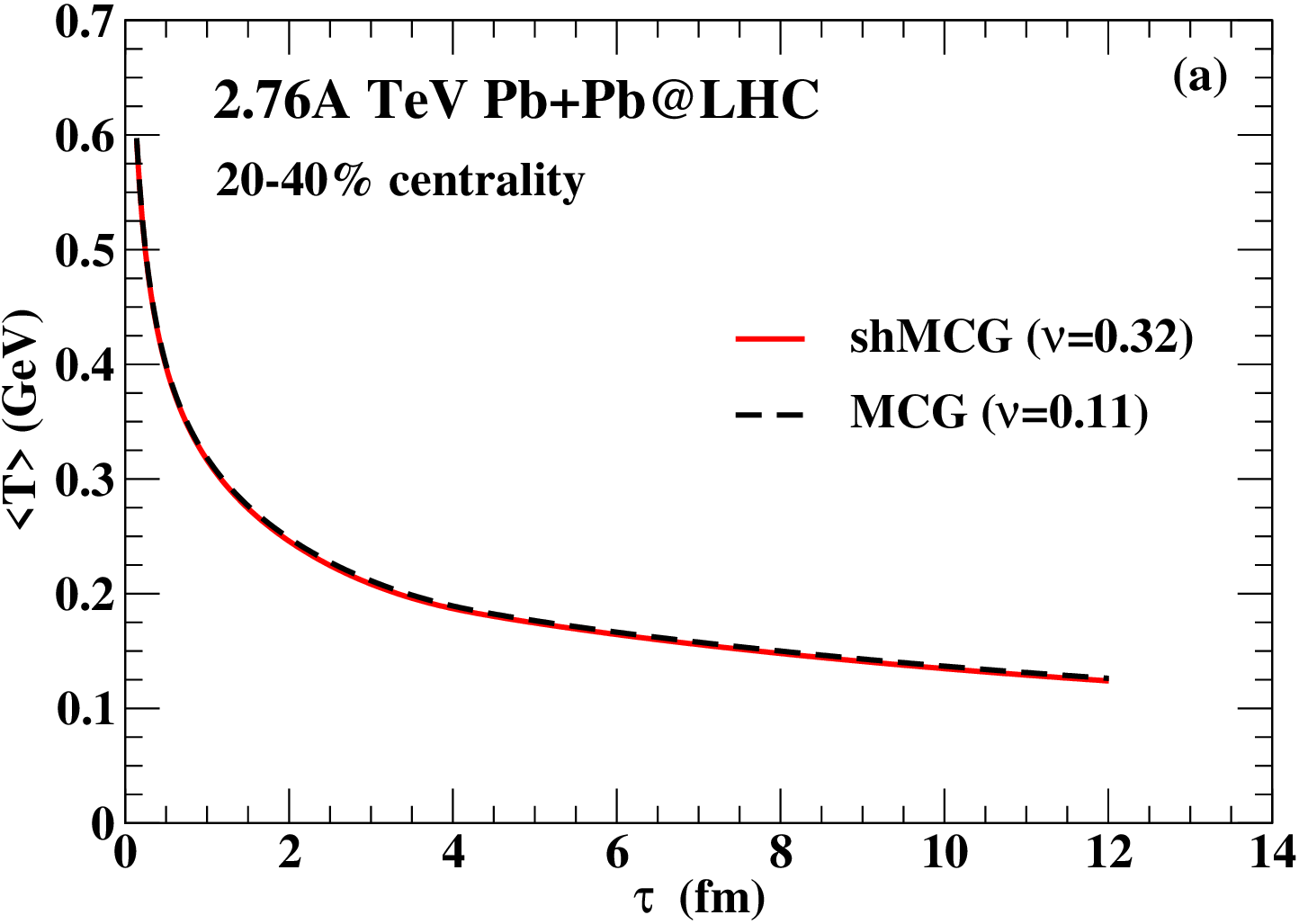}}
\centerline{\includegraphics*[width=8.1cm,clip=true]{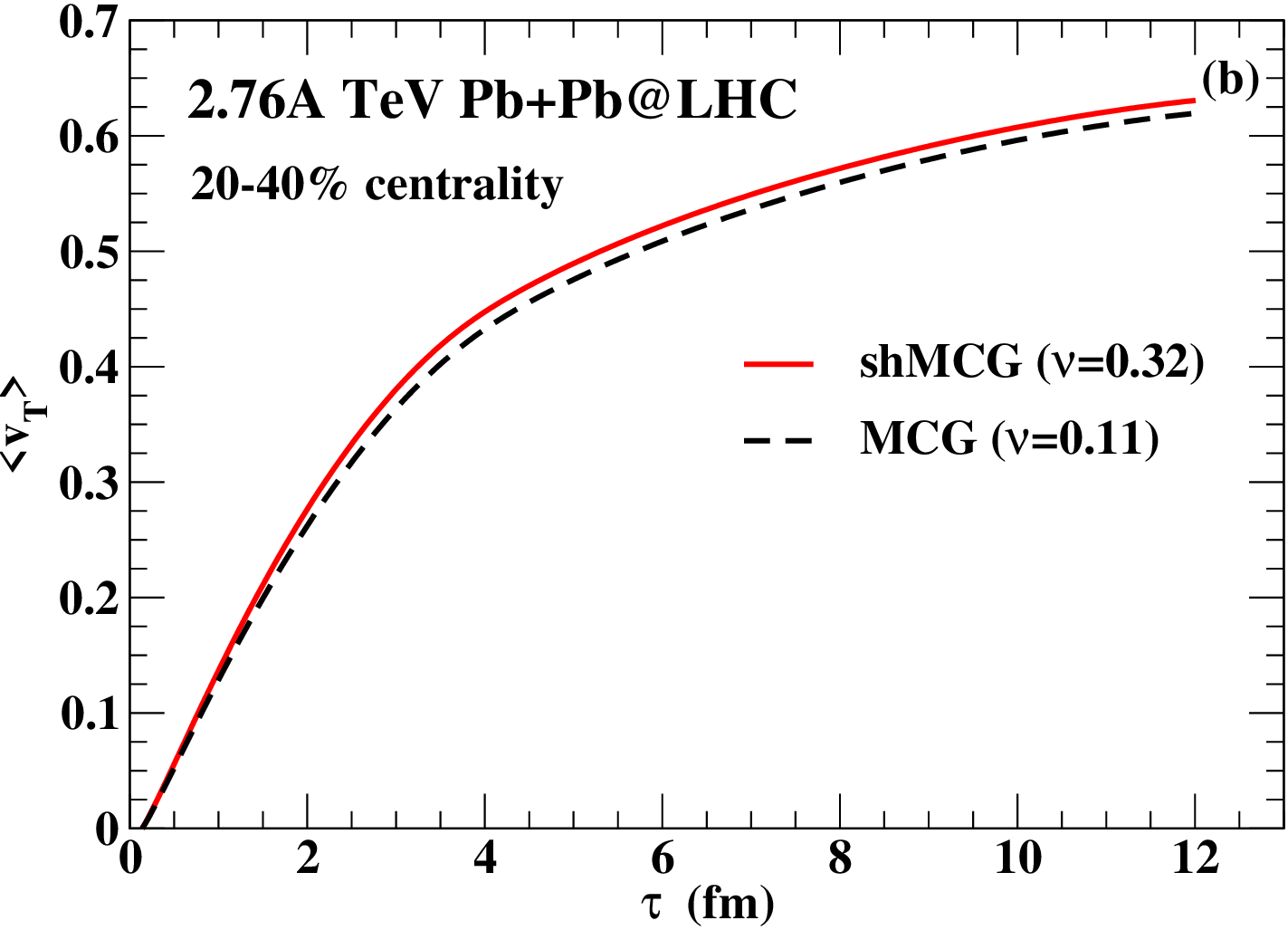}}
\caption{(Color online) (a) The time evolution of temperature and (b) transverse velocity
estimated for LHC energy with two types of initial conditions (see text).   
}
\label{temp_lhc}
\end{figure}

\begin{figure}
\centerline{\includegraphics*[width=8.1cm,clip=true]{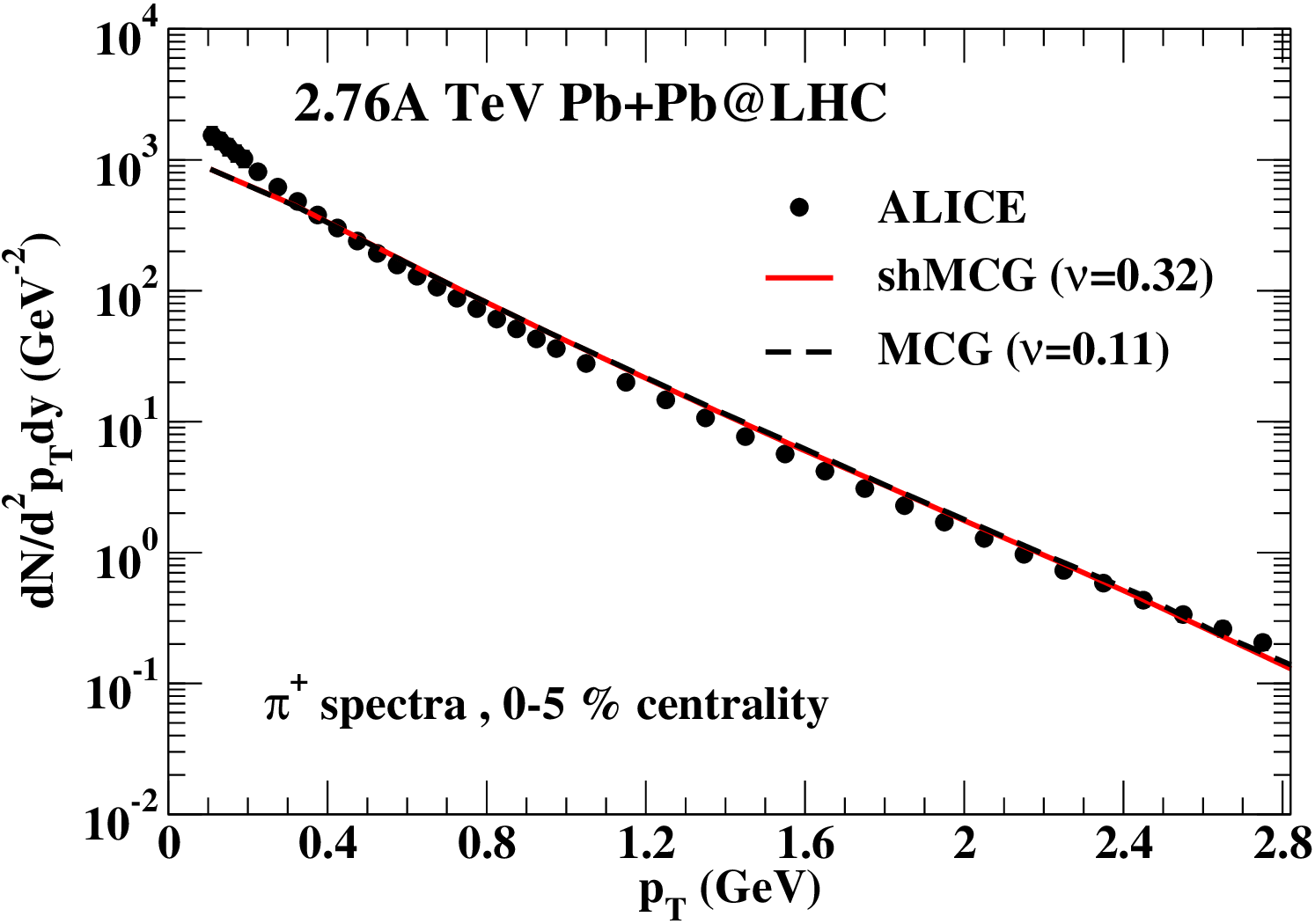}}
\caption{(Color online) The  thermal pion spectra evaluated for 
 LHC with two types of initial conditions (see text) and comparison with ALICE data~\cite{Abelev:2013cb}.   
}
\label{pion_lhc}
\end{figure}

We take $\nu=$0.11, $K=192 \text{ fm}^{-1}$ for the 
MCG initial state and $\nu=$0.32, $K=223 \text{ fm}^{-1}$ 
for the shMCG initial condition to reproduce the same charged hadron 
multiplicity at LHC.
In Fig.~\ref{temp_lhc} we display the variation of average temperature and transverse 
velocity with $\tau$ for LHC energy. 
Although, the results are qualitatively similar to RHIC, however, 
quantitatively the values of $\langle T\rangle$ and $\langle v_T\rangle$ are larger at LHC because
of the larger initial temperature and pressure of the system created in Pb+Pb collisions.

The $\pi^+$ spectra for LHC collision condition have been  evaluated 
at the freeze-out surface and the results are compared with the experimental data~\cite{Abelev:2013cb}. The 
spectra from MCG and shMCG are found to be close to each other (Fig.~\ref{pion_lhc}). 

\begin{figure}
\centerline{\includegraphics*[width=8.1cm,clip=true]{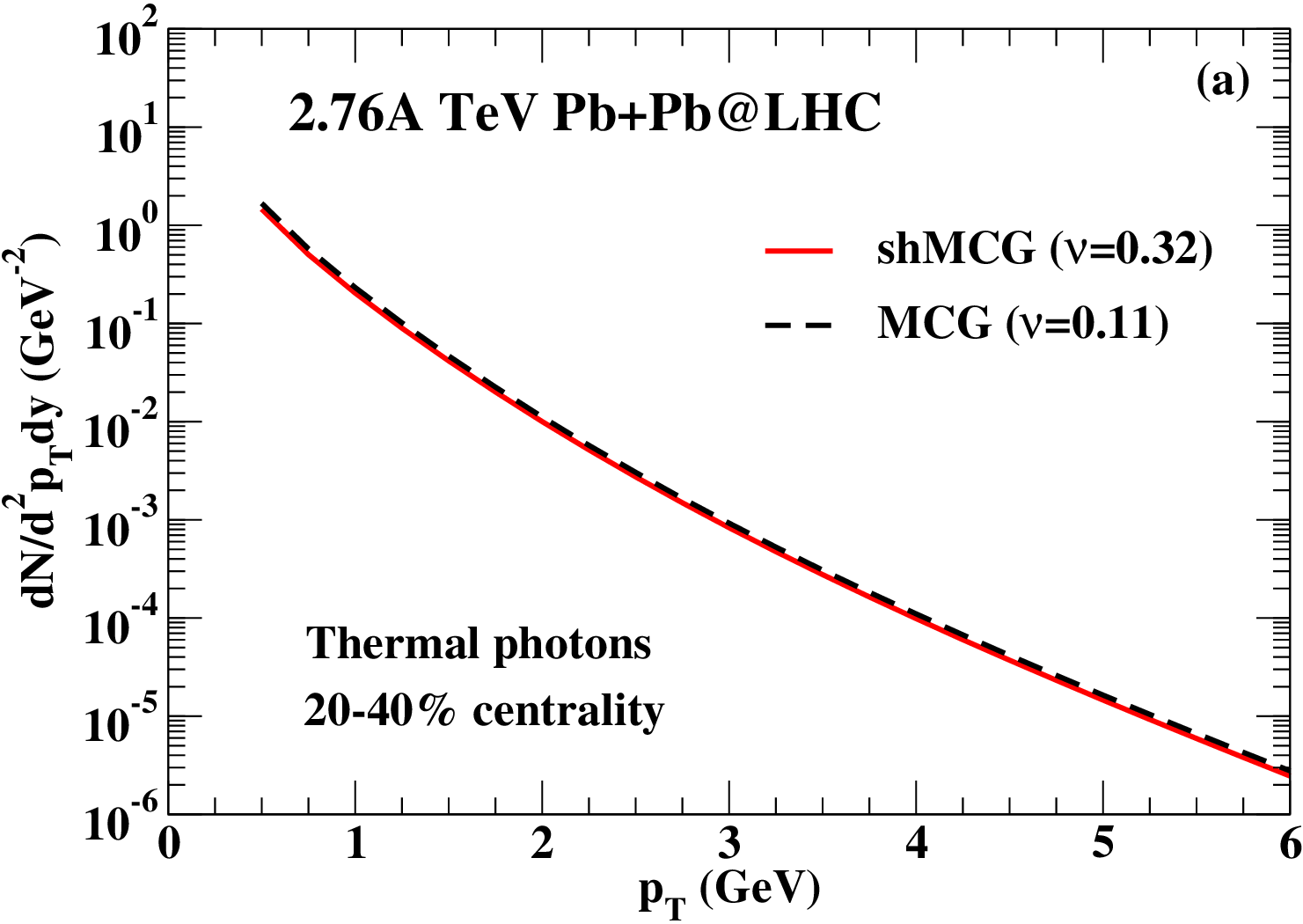}}
\centerline{\includegraphics*[width=8.1cm,clip=true]{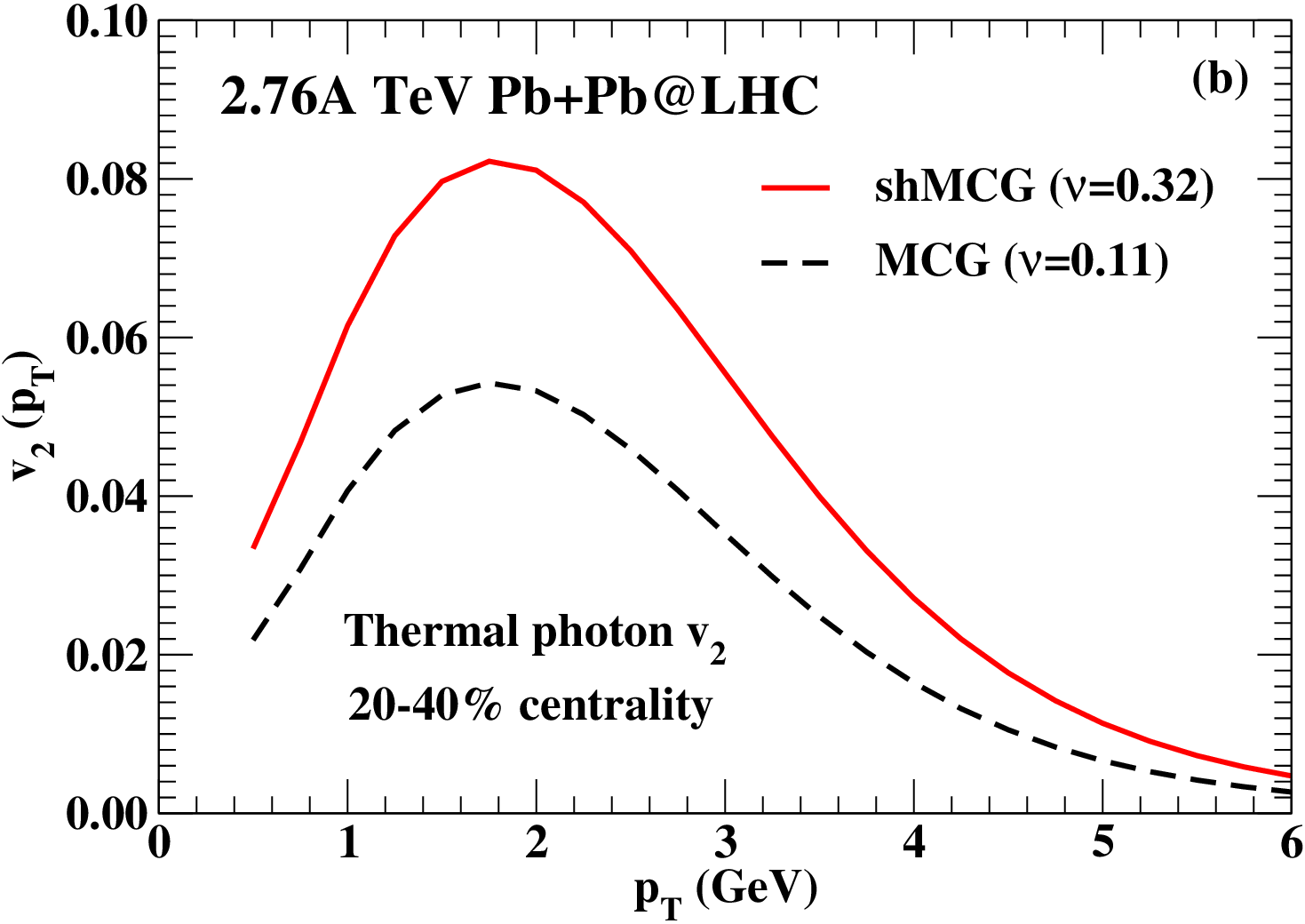}}
\caption{(Color online) (a) The $p_T$ spectra and (b) $v_2$ of thermal photons
for LHC energy with two types of initial conditions (see text).   
}
\label{specv2_lhc}
\end{figure}

Finally, in Fig.~\ref{specv2_lhc} the transverse momentum spectra and 
$v_2$ of photons have been depicted for LHC collision condition. The effects
of shadowing do not show up in the $p_T$ spectra of photons (Fig.~\ref{specv2_lhc}(a)).
However, the elliptic flow (Fig.~\ref{specv2_lhc}(b)) is about 50\% enhanced at the peak value in shMCG as compared to MCG
for reasons explained above.
\begin{figure}
\centerline{\includegraphics*[width=8.1cm,clip=true]{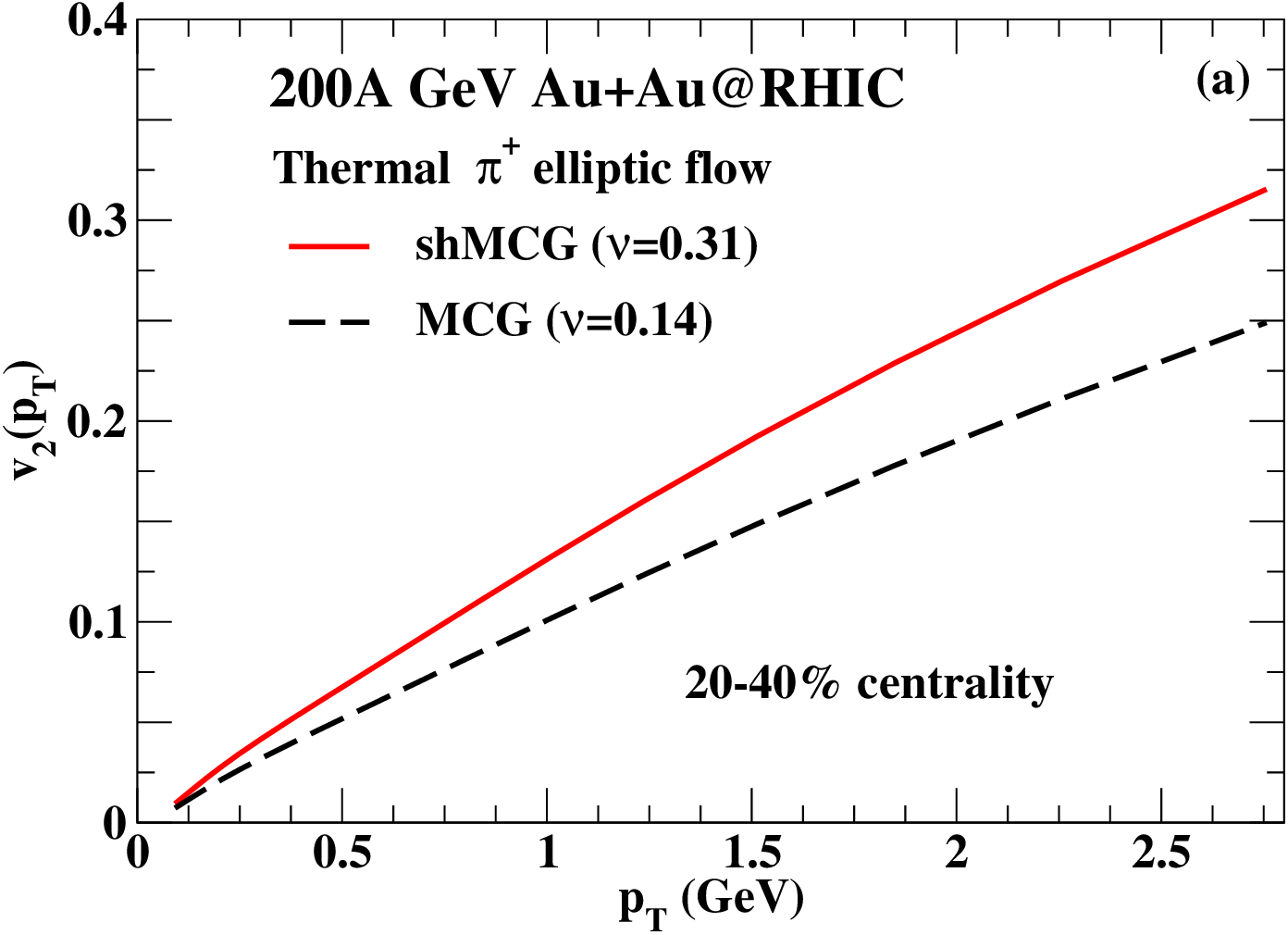}}
\centerline{\includegraphics*[width=8.1cm,clip=true]{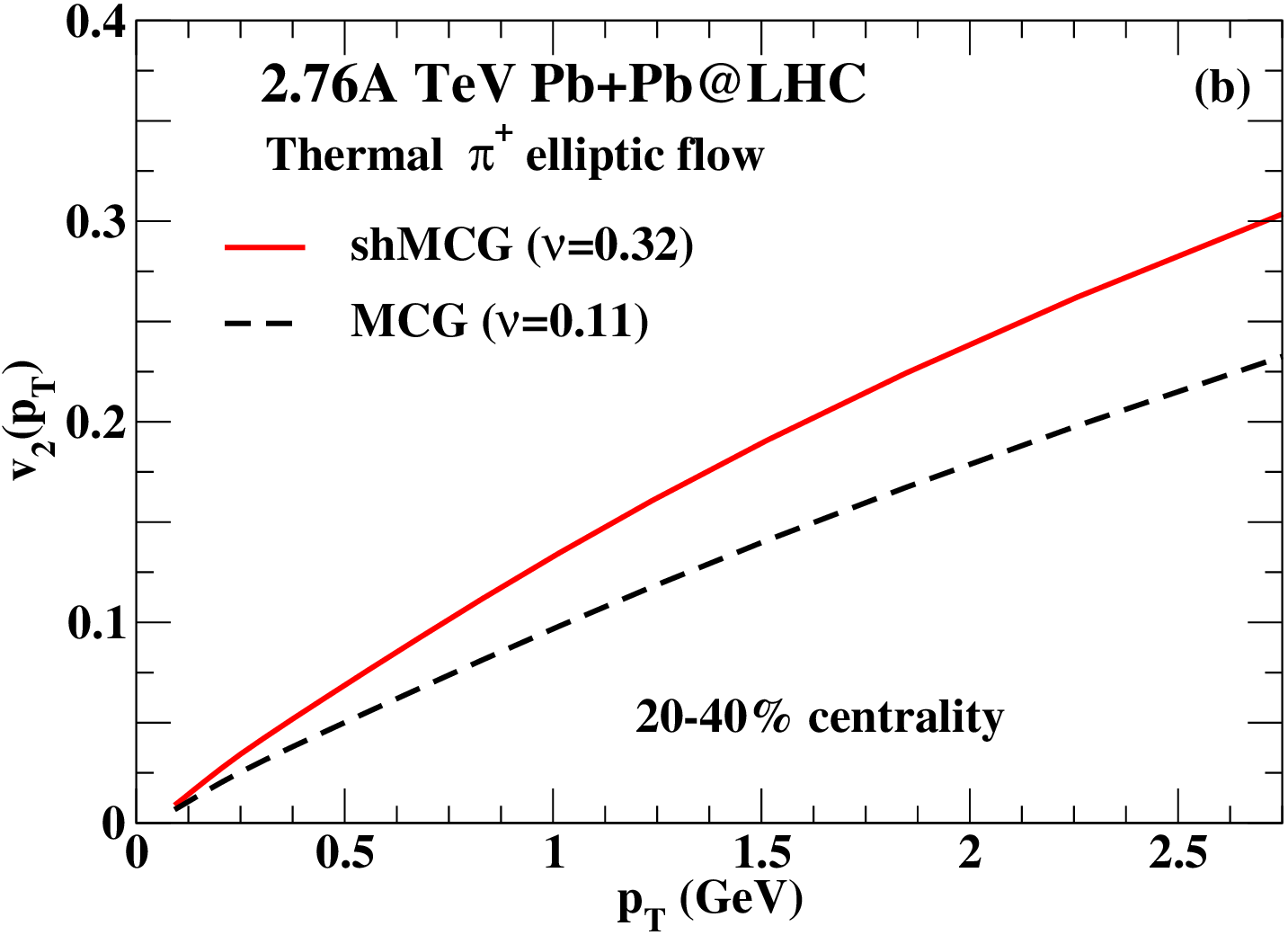}}
\caption{(Color online) (a) The  $v_2$ of thermal pions for RHIC (upper panel)
and (b) LHC (lower panel) initial conditions (see text).   
}
\label{pion_v2_rhic_lhc}
\end{figure}

In order to get some idea about the effect of shadowing on hadronic observables, we calculate the  differential 
elliptic flow of $\pi^+$ for RHIC (Fig.~\ref{pion_v2_rhic_lhc}(a)) and LHC (Fig.~\ref{pion_v2_rhic_lhc}(b)) 
collision conditions with (solid) and without (dashed) shadowing effects.  We notice from the results displayed that 
the  $v_2$ of $\pi^+$ is enhanced by about 29\% for shMCG initial condition compared to the $v_2$ obtained with MCG 
initial condition at $p_T$ = 1.74 GeV (around  $p_T$ =1.74 GeV, $v_2$ of photons attains maximum) at RHIC energy. 
However, the elliptic flow of photons is about 36\% more in shMCG than MCG initial condition at the same $p_T$
for RHIC energy. Similar behavior is observed at the LHC as well. This is indicative of the fact that the photons 
are able to capture the shadowing effects in the initial condition more effectively than  hadrons. 
Also, it is shown in Ref.~\cite{rupa1} that with event-by-event fluctuating initial condition the elliptic 
flow of photons is substantially enhanced. A detailed study is under progress on the elliptic flow of photons and hadrons 
with shadowing and event-by-event fluctuating initial condition  to estimate the 
net enhancement in the $v_2$ of photons~\cite{shadow_ebye_phot_v2}.

\section{summary and conclusions}
We have considered the effects of nucleon shadowing in the MC Glauber initial condition and calculated thermal photon spectra 
and elliptic flow for 200A GeV Au+Au collision at RHIC and 2.76A TeV Pb+Pb collision at LHC for $20-40\%$ centrality bin.
The initial conditions both for MCG and shMCG are constrained to the same experimentally measured charged hadron multiplicity for RHIC. 
Similar exercise has been repeated for LHC
energies for MCG and shMCG initial conditions. Results without these constraints have also been shown. 
Relativistic hydrodynamic equations in (2+1) dimensions have  been solved with lattice QCD EoS to study the space time 
evolution of the system. We calculate photon spectra and elliptic flow considering both wounded nucleon as well as a 
two component model where the initial energy is taken to be proportional to a linear combination of $N_{\rm coll}$ and 
$N_{\rm part}$.  The results on the $p_T$ spectra of photons and pions both at RHIC and LHC collision conditions 
are found to be insensitive to the shadowing effects if the initial conditions are constrained to 
reproduce the same charged hadron multiplicity. However, the elliptic flow of thermal photons from shMCG initial 
condition is found to be significantly larger compared to the MCG initial condition as
shadowing enhances the asymmetry by decreasing the effective ratio of minor to major axis
of the elliptic overlap zone. Therefore, the shadowing effects have  the potential to affect those 
quantities which depend on the geometric asymmetry of the system formed in HICRE. 
The effects of shadowing is found to be smaller in $v_2$ of hadrons than in the $v_2$ of photons. 
In this calculation, we consider initial state averaged smooth profile only which is expected to provide 
a qualitative picture of the effect of initial state shadowing on photons.
Therefore, a complete calculation considering the initial state shadowing in  
event-by-event fluctuating 
initial condition would be useful to  understand the $v_2$ of photons 
better, as it will then contain the contributions from both  initial fluctuations as well as 
non-spherical geometry of the collision zone.

\section{Acknowledgements}
We  thank C\&I group,  VECC for providing computer facility and PDG is grateful to Department of Atomic
Energy for financial support. Discussions with Dinesh Kumar Srivastava and Snigdha Ghosh are gratefully 
acknowledged.

\end{document}